\documentclass[useAMS,usenatbib,usegraphicx]{mn2e}
\newcommand{\cmsq}{cm$^{-2}$}
\newcommand{\cmt}{cm$^{-3}$}

\newcommand{\zsun}{Z$_\odot$}
\newcommand{\kms}{km s$^{-1}$}
\newcommand{\Kkms}{K km s$^{-1}$}
\newcommand{\Xunits}{cm$^{-2}$ K$^{-1}$ km$^{-1}$ s}
\newcommand{\Wunits}{K km s$^{-1}$}
\newcommand{\X}{$X$}
\newcommand{\Xgal}{$X_{\rm Gal}$}
\newcommand{\Av}{$A_V$}
\newcommand{\CO}{CO}
\newcommand{\NCO}{$N_{\rm CO}$}
\newcommand{\NHt}{$N_{\rm H_2}$}
\newcommand{\Ntot}{$N_{\rm tot}$}
\newcommand{\W}{$W$}
\newcommand{\Ht}{H$_2$}

\newcommand {\apgt} {\ {\raise-.5ex\hbox{$\buildrel>\over\sim$}}\ }
\newcommand {\aplt} {\ {\raise-.5ex\hbox{$\buildrel<\over\sim$}}\ } 

\title[Modeling CO Emission from MCs] {Modeling CO Emission: I. CO as a Column Density Tracer and the X-Factor in Molecular Clouds}
\author[R. Shetty et al.]{Rahul Shetty$^{1}$, Simon C. Glover$^{1}$, Cornelis P. Dullemond$^2$, Ralf S. Klessen$^{1,3}$\\
$^{1}$Zentrum f\"ur Astronomie der Universit\"at Heidelberg, Institut f\"ur Theoretische Astrophysik, Albert-Ueberle-Str. 2, 69120 Heidelberg, Germany \\
$^{2}$ Max Planck Institut f\"ur Astronomie, K\"onigstuhl 17, 69117 Heidelberg, Germany\\
$^{3}$ Kavli Institute for Particle Astrophysics and Cosmology, Stanford University, Menlo Park, CA 94025, USA}

\begin{document}

\date{Accepted 2010 November 8. Received 2010 November 8; in original form 2010 
August 10}

\pagerange{\pageref{firstpage}--\pageref{lastpage}} \pubyear{2010}

\maketitle

\label{firstpage}

\begin{abstract}
Theoretical and observational investigations have indicated that the
abundance of carbon monoxide (CO) is very sensitive to intrinsic
properties of the gaseous medium, such as density, metallicity, and
the background radiation field. CO observations are often employed to
study the properties of molecular clouds (MCs), such as mass,
morphology, and kinematics.  It is thus important to understand how
well CO traces the total mass, which in MCs is predominantly due to
molecular hydrogen (\Ht).  Recent hydrodynamic simulations by Glover
\& Mac~Low have explicitly followed the formation and destruction of
molecules in model MCs under varying conditions.  These models have
confirmed that CO formation strongly depends on the cloud properties.
Conversely, the formation of \Ht\ is primarily determined by the
amount of time available for its formation.  We apply radiative
transfer calculations to these MC models in order to investigate the
properties of \CO\ line emission.  We focus on integrated CO (J=1-0)
intensities emerging from individual clouds, including its
relationship to the total, \Ht, and CO column densities, as well as
the ``\X\ factor,'' the ratio of \Ht\ column density to CO intensity.
Models with high \CO\ abundances have a threshold \CO\ intensity
$\approx$65 \Kkms\ at sufficiently large extinctions (or column
densities).  Clouds with low \CO\ abundances show no such intensity
thresholds.  The distribution of total and \Ht\ column densities are
well described as log-normal functions, though the distributions of CO
intensities and column densities are usually not log-normal.  In
general, the probability distribution functions of the integrated
intensity do not follow the distribution functions of CO column
densities.  In the model with Milky Way-like conditions, the
\X\ factor is in agreement with the near constant value determined
from observations.  In clouds with lower metallicity, lower density,
or a higher background UV radiation field, the CO abundances are in
general lower, and hence the \X\ factor can vary appreciably -
sometimes by up to 4 orders of magnitude.  In models with high
densities, the CO line is fully saturated, so that the \X\ factor is
directly proportional to the molecular column density.
\end{abstract}


\section{Introduction}\label{introsec}

Since stars form almost exclusively in clouds which are predominantly
molecular, much effort in current star formation research is focused
on understanding the physics and chemistry of molecular clouds (MCs).
Though MCs are primarily composed of molecular hydrogen (\Ht), due to
the lack of a dipole moment and the unsuitable conditions within MCs
to excite its rotational transitions, \Ht\ is difficult to observe
directly.  Helium, accounting for $\approx 10\%$ of the cloud mass, is
also very difficult to detect.  Carbon monoxide (CO), which is the
second most abundant molecular species in MCs, has a dipole moment
with rotational transitions that are easily excited at typical MC
temperatures (10 - 100 K) and densities (\apgt 100 \cmt).  CO
observations are thus often employed to investigate MC properties.

The observed CO intensity, $I$(CO), which is often expressed as a
integrated ``brightness temperature'' $W_{\rm CO}$ (hereafter $W$), is
considered to be a good tracer of the column density of molecular
hydrogen \NHt.  Namely, \NHt\ is estimated from CO observations
through a constant ``$X$ factor'' \citep[e.g.][]{Dickman78}:
\begin{equation}
X=\frac{N_{\rm H_2}}{W} \, ({\rm cm}^{-2}\,{\rm K}^{-1} \, {\rm km}^{-1} \, {\rm s}).
\label{Xfac}
\end{equation}
CO observations of Galactic clouds have resulted in estimates of
\X\ $\approx$ few $ \times 10^{20}$ \Xunits\ \citep[hereafter \Xgal,
  e.g.][]{ Solomonetal87,Young&Scoville91,Dameetal01}.  Observations
of diffuse gas in the Galaxy have also resulted in similar estimates
of the \X\ factor \citep[e.g.][]{Polketal88, Lisztetal10}.

However, extragalactic observations of systems with different physical
characteristics, such as metallicity or background UV radiation, have
found variations in the \X\ factor.  Interestingly, observational
investigations employing different methodologies have resulted in
vastly discrepant estimates of the \X\ factor.  For example, for the
nearby Small Magellanic Cloud (SMC), \citet{Bolattoetal08} measure a
value $\approx$\Xgal, assuming virialized clouds and using the CO
linewidths to estimate cloud masses, and thereby \NHt.  Independent
dust and gas based observations, on the other hand, suggest extended
regions containing molecular material with little or no CO emission,
resulting in an \X\ factor of up to $\sim$100\Xgal\ \citep{Israel97,
  Rubioetal04, Leroyetal07, Leroyetal09}.

The $^{12}$CO (J=1-0)\footnote{We will hereafter refer to $^{12}$CO
  simply as CO.} line is optically thick in most molecular clouds, and so
\CO\ (J=1-0)
observations are known not to provide direct information about the
total \CO\ mass or column density \NCO.  Observations have shown a
saturation of \CO\ intensities at sufficiently high extinctions
\citep[e.g.][]{Lombardietal06, Pinedaetal08}.  Consequently, only
\CO\ intensities below the saturation threshold are considered in
evaluations of the \X\ factor.  One of the goals of this work is
therefore to assess how well \CO\ observations can trace the true CO
distribution within molecular clouds.

Since the formation of CO is sensitive to the amount of carbon and
oxygen available in the ISM, CO abundances are expected to be lower in
lower metallicity systems \citep{Maloney&Black88, Israel97}.
Additionally, the strength of the background ultraviolet (UV)
radiation field, responsible for photodissociating molecules, also
plays a role in regulating the CO abundances
\citep{vanDishoeck&Black88}.  These processes should in turn lead to
variations in the \X\ factor, with a larger value in metal poor
systems and/or where the UV radiation is higher than in the Milky Way.
Indeed, when independent measures of MC masses are combined with CO
observations, the \X\ factor has been found to be larger in low
metallicity external galaxies, such as the LMC and SMC
\citep{Israeletal86, Israel97, Leroyetal09}.

As turbulence is now considered an important aspect of the dynamics in
molecular clouds and star formation \citep[][and references
  therein]{MacLow&Klessen04,Ballesteros-Paredesetal2007,
  McKee&Ostriker07}, its role in influencing molecule formation needs
to be understood.  \citet{Glover&MacLow07II} \citep[see
  also][]{Glover&MacLow07I}, showed that the formation timescale of
\Ht\ is only a few Myr in turbulent MC models with densities
comparable to those in observed clouds.  Expanding on this work,
through implementation of more extensive chemistry, we can now follow
the formation of CO in models of turbulent molecular clouds \citep[][
  hereafter Paper I]{Gloveretal10}.  Subsequently,
\citet{Glover&MacLow10}, hereafter Paper II, analysed the global
properties of the \X\ factor in different MC models, using spatially
averaged quantities.  They found that the mean extinction, or total
gas column density, primarily determines the \X\ factor.

Here, we explore the observational consequences of the variation in CO
abundances which arise due to differences in MC properties described
in Papers I-II.  Our main goal is to understand the impact of CO
abundance variations within individual MCs on the emerging integrated
CO (J=1-0) intensity.  We consider a suite of models with different
conditions, namely metallicity, density, and background UV radiation
field, representing various environments.  In the next section we
discuss our modeling method, including a brief overview of the
magnetohydrodynamic models with chemistry, and a description of the
radiative transfer calculations.  In Section \ref{resultssec} we
present our results of the comparison of CO intensities with \Ht\ and
\CO\ column densities, and characteristics of the the \X\ factor.  We
discuss our results and compare them to observational investigations
in Section \ref{discsec}.  We conclude with a summary in Section
\ref{sumsec}.

\section{Modeling Method}\label{methosec}

To carry out our investigation of CO emission from molecular clouds,
we apply line radiative transfer calculations to magneto-hydrodynamic
(MHD) models of MCs.  In this section, we give a brief description of
the MHD models, and focus on the radiative transfer calculations.  We
refer the reader to Papers I and II for a more extensive description
of the MHD and chemical modeling method, as well as the analysis of
various simulation runs.

\subsection{Modeling Molecular Clouds: MHD and Chemistry}

The simulations of the model MCs track the evolution of gas using a
modified version of the ZEUS-MP MHD code
\citep{Norman00,StoneNorman92I, StoneNorman92II}.  Gas with initially
uniform density in a 20 pc box with periodic boundary conditions is
driven with a turbulent velocity field, with uniform power between
wavenumbers $1 \le k \le 2$.  Gas self-gravity, which would cause
sufficiently overdense regions to collapse, is not considered in the
calculation.  The simulation includes magnetic fields, with initial
field lines oriented parallel to the $\hat{z}$-axis with a field
strength of 1.95 $\mu$G.  The gas has an initial temperature of 60 K,
but quickly settles to thermal equilibrium.

To track the chemical evolution of the gas, which has constant
metallicity, a treatment of hydrogen, oxygen, and carbon chemistry is
included in the numerical algorithm.  This consists of 218 reactions
of 32 chemical species, which are coupled with the thermodynamics.  A
background UV radiation field, which is responsible for
photodissociation, is treated through the six-ray approximation as
described in \citet{Glover&MacLow07I}.  In our investigation of
emission from CO molecules, we consider one snapshot of these
simulations at a late time $>$ 5 Myr, after which the simulation has
reached a statistically steady state.

\subsubsection{GMC Model Parameters}

The numerical modeling described above follows the combined effects of
turbulence, magnetic fields, thermodynamics, and chemical evolution as
structures such as filaments and dense cores typically found in MCs
forms out of the gas (see Paper I for more details).  In our
investigation of CO emission emerging from the MC simulations, we
consider a suit of models designed to represent various astrophysical
environments.  The relevant parameters of each model are listed in
Table \ref{exptab}.  Column 1 shows the name of each run.  The main
user defined parameters are the initial density $n_0$, metallicity
$Z$, and background UV radiation field $G_0$, indicated in Columns
2-4, respectively.  We assume that the dust-to-gas ratio is directly
proportional to the gas-phase metallicity and do not vary these quantities
independently. Column 5 shows the numerical resolution; as noted
in Paper II, simulations with identical initial conditions but
resolutions of 256$^3$ and 128$^3$ produce very similar
results.\footnote{We also find few differences in the results from
  the radiative transfer calculations applied on those models with
  different resolutions but otherwise identical initial conditions.}
The last column lists the representative environment corresponding to
the simulated cloud: a high density cloud found in the galactic center
(n1000), a typical Milky Way cloud \citep[n300,][]{Ferriere01}, clouds
in low metallicity systems like the LMC or SMC (n300-Z03 and
n300-Z01), a low density cloud in a dwarf galaxy (n100), and clouds in
weak and strong starbursts (n300-UV10 and n300-UV100, respectively).

\begin{table*}
 \centering
 \begin{minipage}{140mm}
  \caption{List of Simulations}
  \begin{tabular}{cccccc}
  \hline
  \hline
  ID  & $n_0$ (cm$^{-3}$) & Z & G$_0$ (2.7$\times 10^{-3}$ erg cm$^{-2}$ s$^{-1}$ ) & Resolution & Representative Environment \\
 \hline
n1000  & 1000 & 1.0 & 1.0 & 128$^3$ & high-density cloud (galactic center) \\
n300  & 300 & 1.0 & 1.0 & 256$^3$ & Milky Way cloud \\
n300-Z03  & 300 & 0.3 & 1.0 & 256$^3$ & LMC/SMC cloud \\
n300-Z01  & 300 & 0.1 & 1.0 & 256$^3$ & LMC/SMC cloud\\
n100  & 100 & 1.0 & 1.0 & 256$^3$ & dwarf galaxy cloud\\
n300-UV10  & 300 & 1.0 & 10.0 & 128$^3$ & weak starburst \\
n300-UV100  & 300 & 1.0 & 100.0 & 128$^3$ & strong starburst \\
\hline
\end{tabular}
 \label{exptab}
\end{minipage}
\end{table*}

\subsection{Radiative Transfer Method}

\subsubsection{Radiative Transfer Overview}

We use the radiative transfer code RADMC-3D (Dullemond et al. in
preparation) to model the molecular line emission of the MHD MC
models.  RADMC-3D is a 3-dimensional code that performs dust and/or
line radiative transfer on Cartesian or spherical grids (including
adaptive mesh refinement).  In this work, since we are only interested
in CO molecular line emission along a chosen direction, we use the ray
tracing capability of RADMC-3D.

One of the primary challenges in line radiative transfer is to solve
for the population levels of the molecular (or atomic) species under
consideration.  The occupation of a given (rotational/vibrational)
energy level of a molecule depends on the incident radiation field, as
well as the collision properties (e.g. frequency) with other atoms or
molecules, both of which act as excitation or de-excitation
mechanisms.  In statistical equilibrium, the relative population of
level $i$, $f_i$, is governed by the equation of detailed balance:
\begin{eqnarray}
\sum_{j>i}[{f_jA_{ji}+(f_jB_{ji}-f_iB_{ij})\bar{J}_{ji}}] -  \nonumber \\
\sum_{j<i}[{f_iA_{ij}+(f_iB_{ij}-f_jB_{ji})\bar{J}_{ij} }] + \nonumber \\ 
\sum_{j}[{f_jC_{ji}-f_iC_{ij}}] = 0
\label{poplevel}
\end{eqnarray}
The last summation accounts for collisions, where $C_{ij}$ is the
collisional rate for a transition from level $i$ to level $j$.  The
collisional rate is dependent on the rate coefficient $K_{ij}$ and the
density of the collisional partner $n_{col}$: $C_{ij} = n_{col}
K_{ij}$.  In MCs, the main collisional partner of CO is \Ht.  We use
the rate coefficients for collisional excitation and de-excitations of
CO by \Ht\ tabulated and freely available from the Leiden database
\citep{Schoieretal05}.  These are based on calculations by
\citet{Yangetal10}.  We neglect the effect of collisions with partners
other than \Ht. We justify this by noting that almost all of the CO in
our simulations is found in regions of the gas that are dominated by
\Ht, rather than atomic hydrogen.

The influence of radiation on setting the level populations is
captured by the first two summations in Equation \ref{poplevel}, which
include the mean integrated intensity $\bar{J}_{ij}$ of the radiation
field in the line corresponding to the transition from $i$ to $j$.
The constants
are $A_{ij}$, the Einstein coefficient for spontaneous emission for a
transition from level $i$ to level $j$, $B_{ij}$, the Einstein
coefficient for stimulated emission from $i$ to level $j$, and
$B_{ji}$, the corresponding coefficient for absorption (also given by
the Leiden database).  Equation \ref{poplevel} is coupled with the
equation of radiative transfer
\begin{equation}
\frac{dI_\nu}{d\tau_\nu}=-I_\nu+S_\nu, 
\label{rteqn}
\end{equation}
where $I_\nu$ is the specific intensity, $S_\nu$ the source function,
and $\tau_\nu$ is the optical depth.  The coupling between Equations
\ref{poplevel} and \ref{rteqn} occurs through the dependence of the
source function ($S_\nu = S_{ij}$) on the relative population levels:
\begin{equation}
S_{ij} = \frac{f_i A_{ij}}{f_j B_{ji} - f_i B_{ij}}, 
\label{sourcefunc} 
\end{equation}
as well as the dependence of the $\bar{J}$ on $I$:
\begin{equation}
\bar{J}_{ij} = \frac{1}{4\pi}\int I_{ij} \phi_{ij} d\Omega,
\label{Jbareqn} 
\end{equation}
where the integral is taken over all solid angles $\Omega$.  The
normalized profile function $\phi_{ij}$ determines the emission and
absorbtion probability of a photon with frequency $\nu$, due to a line
with rest frequency $\nu_{ij}$.  For a photon propagating in a direction
${\bf \hat{n}}$ through a medium with velocity ${\bf v}$,
\begin{equation}
\phi_{ij}(\nu) = \frac{c}{a
  \nu_{ij}\sqrt{\pi}}\exp\left(-\frac{c^2(\nu-\nu_{ij}-{\bf v} \cdot {\bf
    \hat{n}}\nu_{ij}/c)^2}{a^2\nu_{ij}^2}\right), 
\label{lineprof} 
\end{equation}
where $c$ is the light speed.  The thermal and microturbulent
broadening of the line is accounted for through
\begin{equation}
a^2 = v_{mtrb}^2+2k_BT/m_{mol}, 
\label{broaden} 
\end{equation}
where $k_B$ is the Boltzmann constant, $m_{mol}$ is the molecular (or
in the case of atomic lines, atomic) mass, $T$ is the temperature, and
$v_{mtrb}$ is the microturbulent velocity of the medium.

As Equations \ref{poplevel} - \ref{Jbareqn} indicate, the amount of
radiation emitted from molecules in a given location is dependent on
the level populations, which themselves depend on the amount of
incident radiation at that location.  Solving this problem numerically
can be computationally expensive.  In certain situations, however,
suitable approximations can be made which considerably reduce the
computational costs.

One approximation is that of local thermodynamic equilibrium (LTE).
In gaseous systems, when collisional processes dominate line emission,
the temperature is the only parameter that is required to calculate
the population levels, through the use of a partition function.  Thus,
in regions with high \Ht\ densities, employing the LTE assumption to
model the CO (J=1-0) line is reasonable.  However, in the MC
environments considered here, much of the volume does not contain high
densities, so the LTE approximation may not be suitable.\footnote{We
  have also performed radiative transfer calculations assuming LTE to
  obtain CO (J=1-0) line intensities for all the GMC models considered
  in this work. As expected, the LTE intensities only differ by a
  factor of a few from the LVG intensities.  Had we only considered
  LTE, our overall (qualitative) conclusions would be unchanged.  The
  precise values of \W\ and the \X\ factor, however, would be
  appreciably different.}

\subsubsection{The Sobolev approximation}

For this study, we have implemented the Sobolev approximation
(\citealt{Sobolev57}; also known as the Large Velocity Gradient, or
LVG, method) into RADMC-3D to solve for the population levels of an
atomic or molecular species.  This method takes advantage of large
spatial variations in velocity, as are present in turbulent molecular
clouds, to define line escape probabilities.  Effectively, the LVG
method provides a solution to the equation of detailed balance
(Eqn. \ref{poplevel}) determined solely by local quantities.

Consider a photon emitted due to a transition from level $i$ to level
$j$.  Beyond a certain distance from the photon's position of
emission, due to the velocity gradient in the medium, the Doppler-shifted 
frequency associated with the $i$ to $j$ transition is
sufficiently different from that of the incident photon.  The photon
cannot interact with matter at this position, or any subsequent
position, and thus propagates freely out of the cloud.  The LVG escape
probability of the photon can be determined from the optical depth
\begin{equation}
  \beta = \frac{1}{\tau}\int_0^\tau e^{-\tau'} d\tau' = \frac{1-e^{-\tau}}{\tau}.
\label{escprob} 
\end{equation}
Other functional forms of the escape probability can be defined for
different scenarios \citep[e.g. slab or spherical symmetry,
  see][]{vanderTaketal07}.  The optical depth is determined by the
population levels, densities, and velocity gradients $dv/dr$ \citep{vanderTaketal07}:
\begin{equation}
\tau = \frac{c^3}{8\pi\nu_{ij}^3}\frac{A_{ij} n}{1.065 <|dv/dr|>}
\left( f_j \frac{g_i}{g_j}- f_i \right)
\label{tau} 
\end{equation}
where $g_i$ is the statistical weight of level $i$, and $n$ is the
total density of the molecular species ($f_i$ is the relative
population, so that $\sum_i f_i = 1$).  We use the mean (absolute)
velocity gradient $<|dv/dr|>$ across the six faces of each grid zone
in the simulation.  The near unity factor in the denominator is a
correction for integration over a Gaussian line profile.\footnote{We
  have verified that our LVG implementation is accurate by comparing
  the derived population levels of various simple models against those
  provided by the RADEX online LVG caculator \citep{vanderTaketal07},
  and that the LVG solution at high densities matches the LTE
  solution.}

Once the optical depth is computed, the escape probability is given by
Equation \ref{escprob}.  The local radiation field is then determined
by
\begin{equation}
\bar{J_{ij}} = S_{ij}(1-\beta)
\label{Jbar} 
\end{equation}
Numerically, given these expressions for $\tau$, $\beta$, $\bar{J}$,
and the local velocity gradient, one can iteratively solve for the
population levels in Equation \ref{poplevel}; for more detailed
descriptions of the LVG method see \citet{Mihalas78} and
\citet{Elitzur92}.

Though the LVG method was originally formulated and developed for
spectral line studies in stellar atmospheres which have smooth
velocity gradients \citep{Sobolev57, Castor70, Mihalas78}, it can also
be applied for any environments with significant velocity gradients.
In fact, \citet{Ossenkopf97} showed that the Sobolov method is a very
good approximation for computing level populations in MCs, where the
velocity gradients are generally not smooth.\footnote{A photon
  determined to escape through an LVG calculation may be absorbed in a
  distant region for a system with a stochastic velocity distribution.
  We do not consider such events, as they are not expected to occur
  frequently in systems that are highly supersonic.}

\subsection{Analysis of CO Emission}

In this work, we focus on how well CO emission can trace the intrinsic
CO column density, \NCO, the \Ht\ column density \NHt, as well as the
total column density of hydrogen nuclei, \Ntot. The intrinsic column
densities can be easily calculated directly from the MHD simulation by
integrating the CO, \Ht\ and hydrogen nuclei volume densities along a
given axis.  Accordingly, for the radiative transfer calculation we
orient the simulation cube such that the CO line is ``observed'' along
the same axis for which the column densities are computed (e.g. along
the $\hat{z}$-axis).  We have verified that our results are not
sensitive to the choice of orientation.

Besides the viewing geometry, the only other user defined parameter
required for the radiation transfer calculations is the microturbulent
velocity $v_{mtrb}$ [see Eqns. \ref{lineprof}-\ref{broaden}].
Extrapolating the observed linewidth-size relationship
\citep[e.g.][]{Larson81} down to the resolution of the MHD simulations
$\sim$ 0.1 pc, appropriate microturbulent velocities are in the range
$\sim$ 0.2 - 0.7 \kms.  We have explored this range in $v_{mtrb}$, and
have found that the results are insensitive to the particular choice
of $v_{mtrb}$.  In the analysis presented here, $v_{mtrb}$ is set to
0.5 \kms.

In our discussion, we will sometimes express \Ntot\ as an extinction
\Av, in order to allow for direct comparison with observational
analyses.  Observers are often required to employ indirect measures of
\Ntot, since \Ht, a major constituent of \Ntot\ in MCs, is difficult
to observe directly.  Extinction measurements provide estimates of the
total amount of dust along the line of sight.  Using a ``reddening
law,'' (and an assumption for the dust-to-gas ratio), the total
gaseous column follows directly from the amount of extinction
\Av\ \citep{Bohlinetal78}.  In many Galactic molecular clouds, nearly
all the hydrogen is molecular, so \NHt\ is directly proportional to
\Av.  In other environments, however, such as those with lower
metallicity or higher UV fields (e.g. Models n300-Z01 and n300-UV2),
there may be significant amounts of atomic hydrogen, so \Ntot, and
hence the extinction, is dependent on both the molecular and atomic
column densities. To allow for straightforward comparison between
models and observations, we use a simple conversion between \Ntot\ and
\Av:
\begin{equation}
A_{V}=\frac{N_{\rm tot}}{1.87 \times 10^{21} \, {\rm cm}^{-2}} \: \left(\frac{\rm Z}{\rm Z_{\odot}}\right),
\label{extinction}
\end{equation}
where ${\rm Z}$ is the metallicity of the gas. Thus, in the comparison of CO emission with intrinsic cloud
properties, any discussion involving \Av\ can be directly translated
into total column density.

The radiative transfer calculations produce spectral
(position-position-velocity, or PPV) cubes of the CO (J=1-0) line.
The 3D cube indicates the intensity $I_\nu$ in a given frequency or
velocity channel of width $dv$ at each 2D position.  We choose a
sufficiently large range in frequencies such that all (line-of-sight)
velocities in the simulation are detected, so that all emission from
the model MC is ``observed.''

In the comparison of CO intensities with intrinsic column densities,
as well as the analysis of the \X\ factor, the quantity of interest is
the velocity integrated intensity, which is simply the PPV cube
integrated over its velocity axis.  The intensity $I_\nu$, which has
units of erg s$^{-1}$ cm$^{-2}$ Hz$^{-1}$ ster$^{-1}$, can be
expressed as the Planck function evaluated at a ``brightness
temperature'' $T_B$, $B_\nu(T_B)$.  Since the CO (J=1-0) line is
located in the Rayleigh-Jeans part of the spectrum, $I_\nu \propto
T_B$.  The intensity of CO line emission is thus often conveyed in
$T_B$ units.  We follow this convention and express the intensity as a
velocity integrated brightness temperature:
\begin{equation}
W=\frac{1}{2k_B} (c/\nu)^2 \int I_\nu dv \,\, ({\rm K \, km \, s^{-1}}). 
\label{Weqn}
\end{equation}
This integrated intensity is thus a measure of total CO emission along
the line of sight.  \W\ is computed at all positions yielding a 2D
map, which can be used with the 2D map of \NHt\ to obtain the
\X\ factor through Equation \ref{Xfac}.

\section{Results} \label{resultssec}

\subsection{CO Intensities and Column Densities \label{intensitysec}}

\begin{figure*}
\includegraphics[width=175mm]{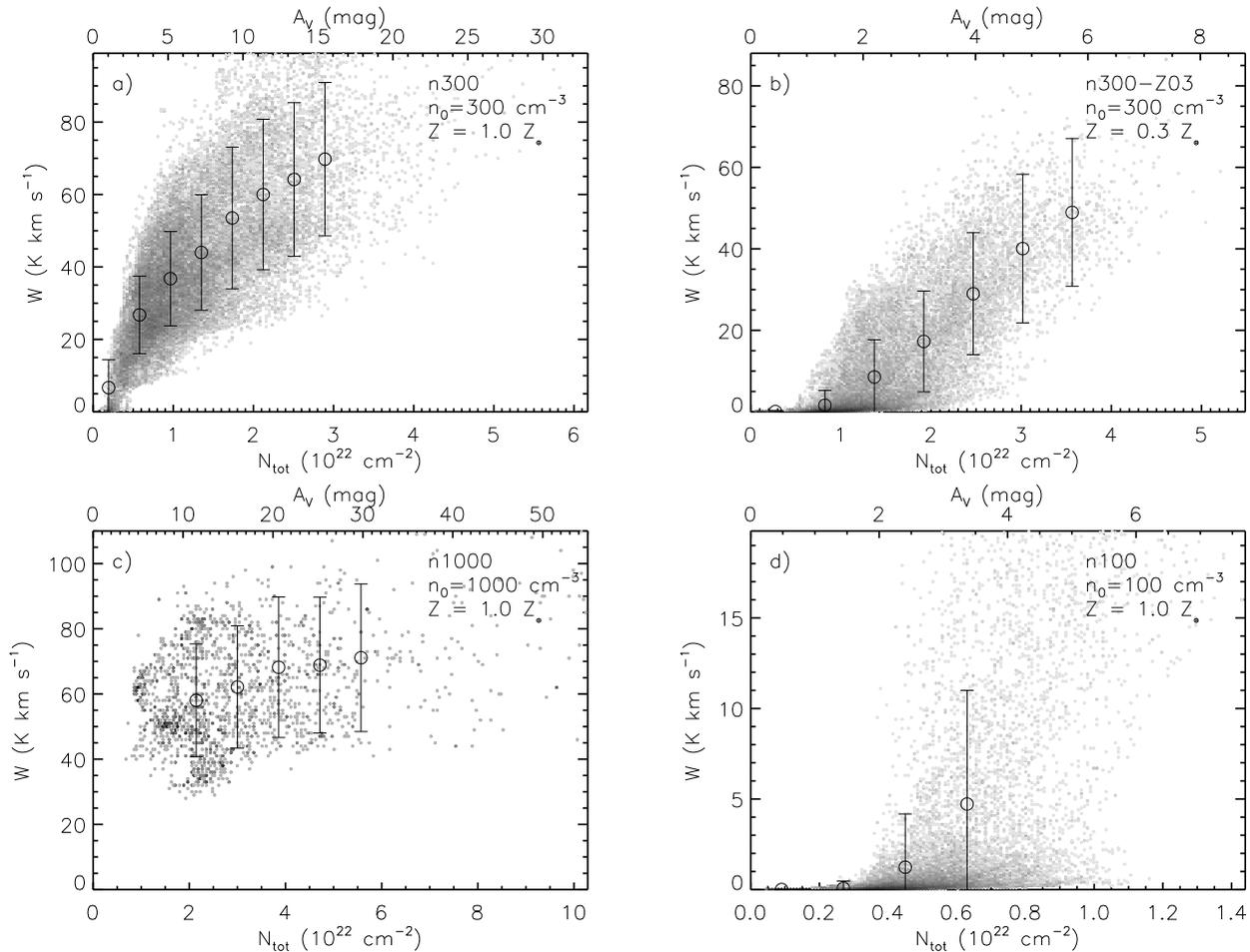}
\caption{Relationship between integrated CO intensity \W\ and total
  column density \Ntot\ (bottom abscissa) or extinction \Av\ (top
  abscissa) for four models: a) n300, b) n300-Z03, c) n1000, and d)
  n100.  The color of each point indicates the frequency of a given
  \Av\ and \W\ pair, with darker points corresponding to higher
  frequencies.  Circles show average \W\ values in \Av\ bins, and the
  error bars indicate 1$\sigma$ deviations.}
\label{wvsav}
\end{figure*}

To begin our investigation of CO emission, we assess the correlation
of velocity integrated \CO\ intensities \W\ with the total column
density \Ntot\ of the model MCs.  Figure \ref{wvsav} shows \W\ as a
function of the \Ntot\ for four of the simulations listed in Table
\ref{exptab}.  The bottom abscissa shows \Ntot, while the top abscissa
shows the corresponding \Av\ (see Eqn. \ref{extinction}).

A number of features are readily apparent in Figure \ref{wvsav}.
First, there is a general trend of increasing intensity with
increasing column density, though the slopes differ between the
various simulations.  For the high-density run (n1000) shown in Figure
\ref{wvsav}c, the \CO\ intensities do not demonstrate any clear trend
with increasing \Av.  Rather, the vast majority of the intensities
reach a threshold value of $\sim$65 \Wunits.  This saturation of
\CO\ intensities is expected to occur at high densities since the
\CO\ line becomes optically thick.  Saturation is also found at the
highest extinctions in the Milky Way simulation (n300, Fig
\ref{wvsav}a).  Indeed, saturation at high CO intensities has been
observed in MCs in the solar neighborhood
\citep[e.g.][]{Lombardietal06,Pinedaetal08}.  Though model n300
qualitatively reproduces the observed trends of increasing CO
intensity with increasing density, up to a threshold value, a detailed
quantitive comparison is not appropriate here; the precise slope,
scatter, minimum, and threshold intensity are dependent on additional
physics not included in our models, such as additional heating due to
stars, outflows, and supernovae.

Figure \ref{wvsav}b and d show that for clouds with lower
metallicities (Model n300-Z03) or densities (Model n100), there is a
wider distribution of \CO\ intensities at low extinctions.  The
saturation of the CO line is not easily evident compared to Model
n1000.  Since CO is easily photo-dissociated in low metallicity or low
density systems due to insufficient (dust and self-) shielding, the CO
column densities do not reach high enough values for the saturation of
the \CO\ line to occur.  Observations have indeed shown that the
\CO\ intensities are generally lower in extragalactic systems with
lower metallicities than that of the Milky Way.  Further, there is no
evidence of saturation in the CO line from diffuse sources such as the
Polaris Flare, LMC and SMC
\citep[e.g.][]{Heithausen93,Israel97,Leroyetal09}.

\begin{figure*}
 \centering
\includegraphics[width=150mm]{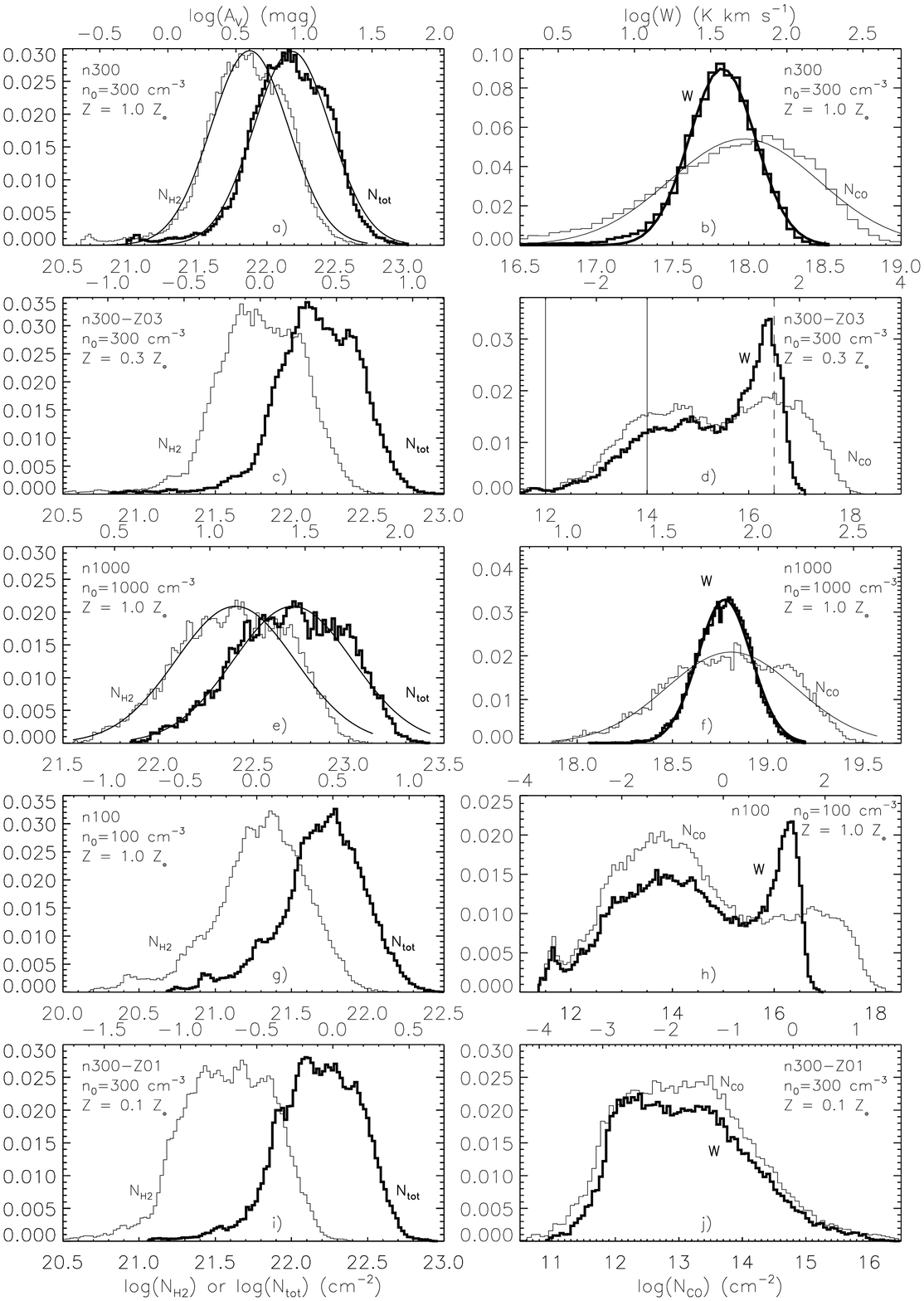}
\caption{PDFs of {\it left:} molecular hydrogen column density
  \NHt\ (thin) and total column density \Ntot\ (thick), {\it right:}
  CO column density \NCO\ (thin) and integrated CO intensity
  \W\ (thick) from five models.  In the panels on the left, the bottom
  abscissas indicate column density, and top abscissas indicate total
  extinction.  In the panels on the right, the bottom abscissas
  indicate \NCO\ values, and top abscissas indicate \W\ values.  For
  Model n300-Z03 (c-d), solid lines mark log(\NCO)=12, 14 and
  log(\W)=-3, -1 and dashed lines mark log(\NCO)=16.5 and log(\W)=1.5,
  for comparison with Figure \ref{compfigs} (see text).  Best fit
  Gaussians are overplotted for Models n300 (a-b) and n1000 (e-f).
  Model parameters are indicated within each plot.}
\label{pdfpans}
\end{figure*}

Figure \ref{pdfpans} shows the probability distribution functions
(PDFs) of \Ntot\ (or \Av), \NHt, \NCO, and \W\ for five models.  The
\Ntot\ PDFs are offset to larger values compared with the \NHt\ PDFs.
The offset for the Milky Way (n300) and the high density cloud (n1000)
can be wholly attributed to the difference in \Ntot, which is a
measure of the total number of hydrogen nuclei, and \NHt, a measure of
the number of molecular hydrogen nuclei.  For the other models, there
remains an offset even after accounting for the factor of 2 difference
between \Ntot\ and \NHt, because significant amounts hydrogen remains
in atomic form (see Table 2 of Paper II).

The \Ntot\ and \NHt\ PDFs of all simulations can be well described as
log-normal functions.  Log-normal column density PDFs should be
expected from these compressible turbulent simulations
\citep{Vazquez-Semadeni94, Klessen00, Ostrikeretal01, Lietal03,
  Federrathetal10}.  The simulation with conditions similar to the
Milky Way (n300) and with the highest density (n1000) produce
\NCO\ and \W\ PDFs which come closest to being lognormal.  In general,
however, the CO column densities and intensities do not show a
log-normal PDF.  We defer the discussion of the best fit Gaussians
shown for these two models to Section \ref{logdisc}.

Another clear trend in Figure \ref{pdfpans} is that the shapes of the
\W\ PDFs can differ significantly from the PDFs of underlying \NCO.
The extent of the scales of \W, shown on the top abscissas of the
right hand plots in Figure \ref{pdfpans}, is equivalent to the extent
of the scales in \NCO, which appear on the bottom abscissas.  At low
densities where gas is optically thin, CO emission should trace the CO
density.  Thus, the absolute scales are chosen such that low \W\ and
\NCO\ values are matched as best as possible.  For models n300 (Figs
\ref{pdfpans}a-b) and n1000 (Figs \ref{pdfpans}e-f) , the \W\ PDFs
have steep increases, resulting in peaks that do not correspond to the
peaks in the \NCO\ PDFs.  The PDFs of the other simulations, n300-Z03,
n300-Z01, and n100, show comparable gradients at low \W\ and \NCO, but
at high values the \W\ PDFs drop more rapidly than \NCO.

In those PDFs showing a steep decline in \W, at high CO column
densities there is little or no corresponding CO emission, even though
there is good correspondence at lower values (due in part to the
choice of the \W\ and \NCO\ axis ranges).  PDFs showing a much steeper
gradient beyond the \W\ peak compared to the \NCO\ peak can be
attributed to saturation of the CO line: as a result of saturation,
regions with very high \NCO\ may have similar intensities to regions
with lower (but still high) \NCO, producing a PDF which appears to
have a ``piled-up'' profile at some high \W.  Such PDF shapes are
clearly evident in the moderately low metallicity (n300-Z03) and very
low density (n100) cloud simulations.  For the simulation with very
low metallicity (n300-Z01) for which the \W\ and \NCO\ PDFs appear to
be correlated, there is no evidence of a saturation feature.

\begin{figure*}
\includegraphics[width=175mm]{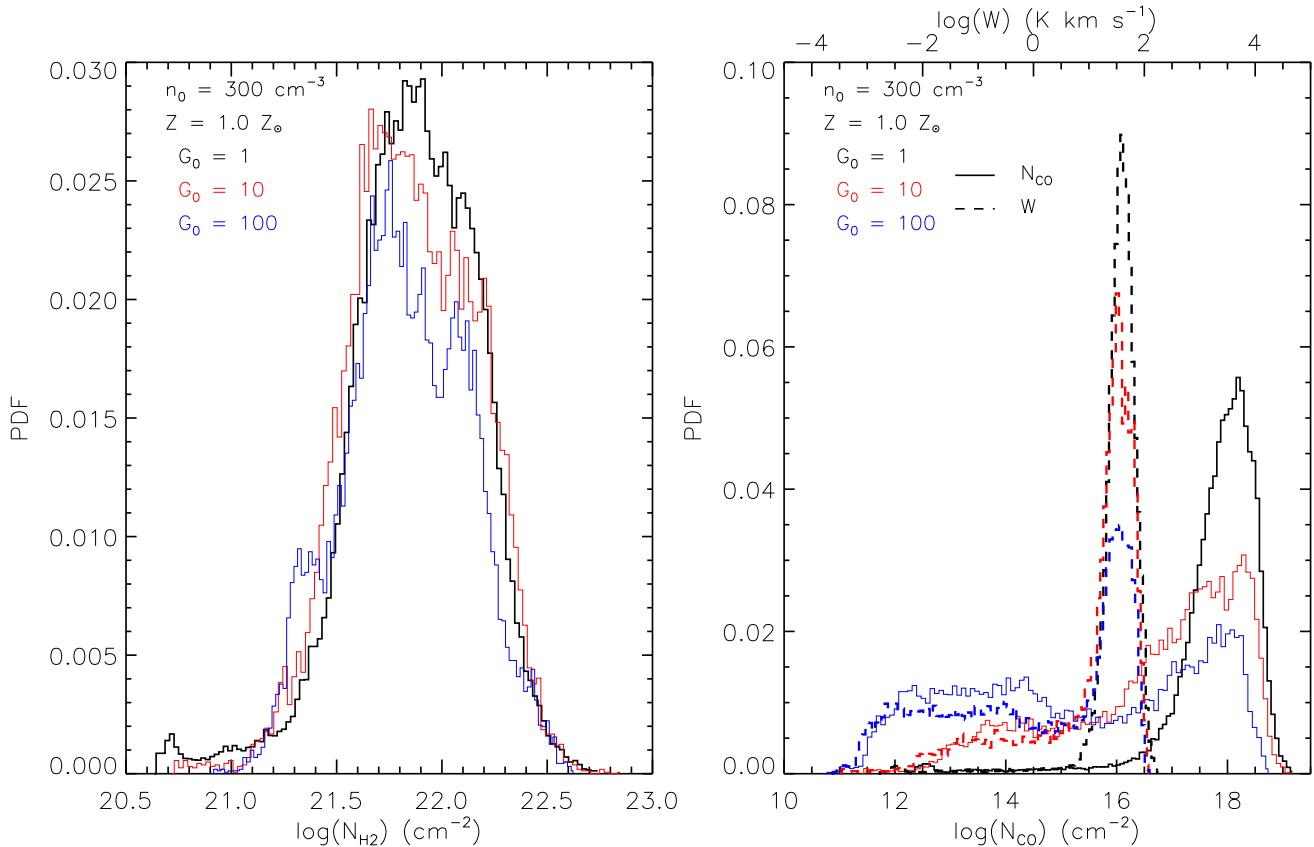}
\caption{PDFs of {\it left:} \NHt, {\it right:} \NCO\ (solid
  thin) and \W\ (dashed thick) from three models with different
  background interstellar radiation fields (indicated within each
  plot).  In the panel on the right, the bottom abscissa indicates
  \NCO\ values, and top abscissa indicates \W\ values.}
\label{highUV}
\end{figure*}

We have shown that there is no simple correlation between \W\ and
\NCO, focusing on models with different metallicities and densities.
Of course, the background radiation field also affects the formation
of CO, due to UV photodissociation \citep[][Paper
  II]{vanDishoeck&Black88}.  Figure \ref{highUV} shows the \NHt, \NCO,
and \W\ histograms from models n300, n300-UV10, and n300-UV100 (see
Table \ref{exptab}).  These models only differ in their background UV
radiation field strengths, with values of G$_0$ = 1, 10, and 100 times
the Galactic estimate (2.7$\times 10^{-3}$ erg cm$^{-2}$ s$^{-1}$).
All simulations have similar log-normal \NHt\ distributions.
Simulations with higher UV fields have more extended \NCO\ PDFs, and
thereby lowered maxima in the \NCO\ probabilities.  This occurs
because stronger UV radiation is able to penetrate further into the
cloud to dissociate CO, resulting in more regions with low CO
densities.  Similarly, the \W\ distributions are more extended for
models with brighter UV backgrounds, with fewer regions with high
\W\ values.  These results simply follow the trend previously
discussed: models with a wide range in CO abundances demonstrate a
lack of correlation between \W\ and \NCO\ PDFs.  The lack of
correlation in the PDFs primarily occur at the highest densities and
intensities (with the exception of n300-Z01, but see last paragraph of
this subsection), though at lower densities gas is optically thin
resulting in a correlation between \W\ and \NCO.  Clouds with lower
metallicity, lower overall density, higher UV background, or a
combination of these factors have much more distributed CO abundances,
resulting in greater discrepancies between the \W\ and \NCO\ PDFs.
Keeping this general trend in mind, in our subsequent analysis we will
focus on the models with G$_0$ = 1.

\begin{figure*}
\includegraphics[width=175mm]{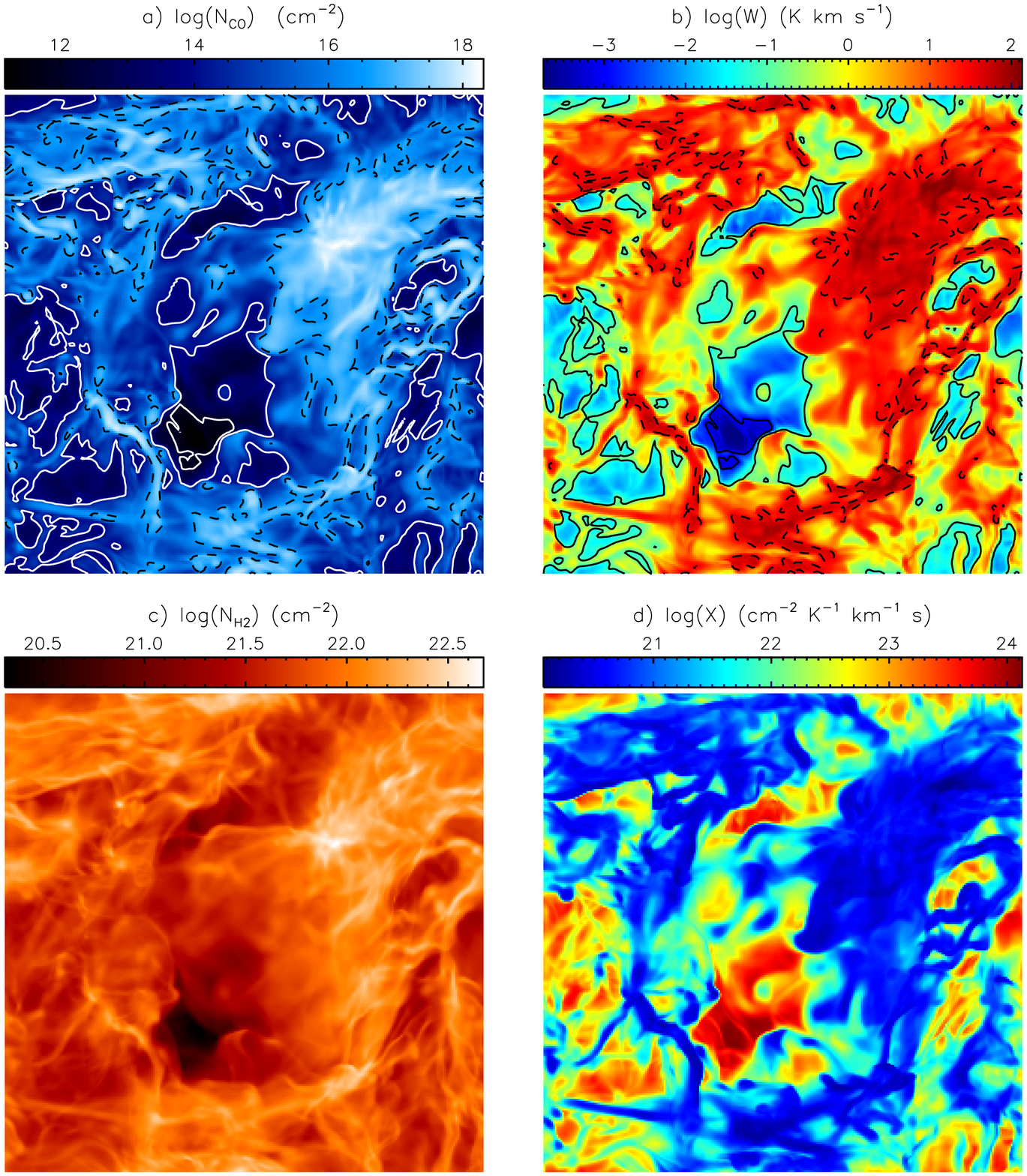}
\caption{Images of a) \NCO, b) \W, c) \NHt, and d) the \X\ factor of
  Model n300-Z03.  Each side has a length of 20 pc.  In a) and b)
  solid contours indicate log(\NCO)=12, 14 and log(\W)=-3, -1; dashed
  contours are log(\NCO)=16.5 and log(\W)=1.5 (see text and
  Fig. \ref{pdfpans}d).}
\label{compfigs}
\end{figure*}

We now turn our attention to the \W\ and \NCO\ properties of
simulation n300-Z03 (Figs. \ref{pdfpans}c-d).  There is a relatively
large spread in \NCO, which also has a clear double peak, at
$\log(N_{\rm CO}) \approx$ 14.5 and 16.5 (with similar probabilities).
Further, though the \W\ PDF appears to correlate well with the
\NCO\ PDF at low CO density (or low intensity) values, the peak in
\W\ does not correlate well with the peak in \NCO.  A closer
inspection of the \NCO\ and \W\ images for Model n300-Z03, shown in
Figure \ref{compfigs}a-b, further reveals the correlation at low
densities, and lack of correlation at high densities.  In the low
intensity range $-3 \leq$ log(W) $\leq \, -$1, marked by solid lines
in Figure \ref{pdfpans}d, an increase in \W\ corresponds to a similar
increase in \NCO\ in the 12 $\leq$ log(\NCO) $\leq$ 14 range.  These
\W\ and \NCO\ values are indicated by solid contours in Figure
\ref{compfigs}, showing excellent correspondence. However, at higher
intensities, \W\ does not faithfully trace the high \NCO\ regions.
The PDF of intensities at log(\W) $\geq$ 1.5, marked by a dashed line
in Figure \ref{pdfpans}d, decreases sharply, whereas log(\NCO)
continues to increase.  The dashed contours in Figure \ref{compfigs}
mark the corresponding regions in the \NCO\ and \W\ maps.  In some of
the filaments (e.g. near the bottom), the contours mark similar
regions.  However, in the most dense regions (e.g. the large
over-dense region in the top right), the area within the \W\ contour
is noticeably smaller than that encompassed in the \NCO\ map.  The
detailed morphology in the region encompassed by the dashed contour
also shows stark differences between the observed CO intensities and
intrinsic CO column densities.  These differences at high densities
are a direct consequence of the optically thick nature of the
\CO\ line.

Similar discrepancies are present in a \NCO\ - \W\ comparison of all
the models.  In some cases, such as model cloud with very low
metallicity, n300-Z01 shown in Figure \ref{pdfpans}j, the apparent
correlation in PDFs at low densities does not translate into a
correlation in the 2D maps.  As stated, the \NCO\ and \W\ PDFs were
plotted with abscissas chosen to best match the PDFs at low densities
and intensities, and so any apparent correlation may only be
coincidental.  

In summary, we have shown that at low densities, CO emission traces
the distribution of CO molecules well, since the CO line is optically
thin.  However, at high densities, the line becomes optically thick,
resulting in an intensity threshold.  Thus, CO intensity does not
neatly trace the highest density gas.  Further, there is generally no
simple correlation in the distribution of CO and \Ht\ molecules, since
\Ht\ can better shelf-shield against photodissociation.  Taken
together, the lack of any general correspondence between \W\ and \NCO,
as well as \NCO\ and \NHt, will certainly affect further analysis of
CO observations, such as attempts at relating \W\ to \NHt.

\subsection{The \X\ Factor \label{xfacsec}}

In Paper II, the \X\ factor (Eq. \ref{Xfac}) from the models
considered in this work was found to vary at low mean extinctions
$\overline{A_V}$ \aplt\ 3.  At higher mean extinctions, the \X\ factor
remained constant.  In that work, the CO intensities were estimated by
assuming local thermodynamic equilibrium (LTE), and by averaging
relevant quantities over whole clouds.  In the previous section, we
demonstrated that \CO\ observations do not neatly trace \CO\ column
density.  A logical expectation from this result is that observed
\CO\ intensities would not be a good tracer of \Ht\ number density,
leading to a variation in the \X\ factor as found in Paper II.  In
this section, we go beyond the analysis of Paper II to investigate the
\X\ factor using the results from the 3D radiative transfer
calculations to estimate \CO\ line intensities.  Here, we are also
able to assess variations in the \X\ factor on a position-by-position
basis in maps of the individual clouds.

\begin{figure*}
\includegraphics[width=175mm]{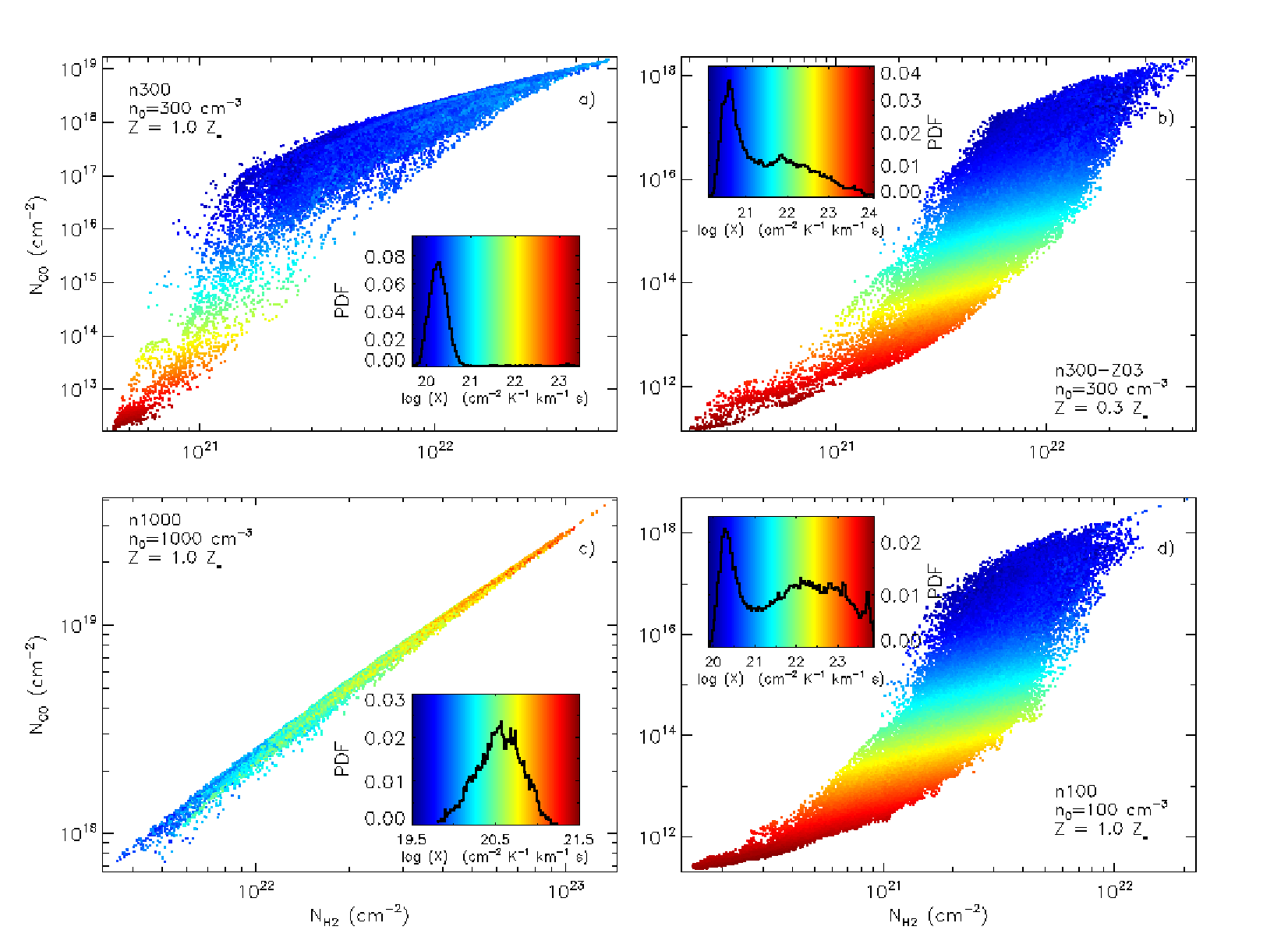}
\caption{\X\ factor for four models. \NCO\ is plotted as a function of
  \NHt. The color of each point indicates the \X\ factor.  Inset
  figures show the color scale and PDF of the \X\ factor.  The
  corresponding maps of \NHt, \NCO, and the \X\ factor from Model
  n300-Z03 are shown in Figure \ref{compfigs}.}
\label{Xfig}
\end{figure*}

Figure \ref{Xfig} shows the \X\ factor from each position in the 2D
integrated intensity images of simulations n300, n300-Z03, n1000, and
n100, along with the PDF of the \X\ factor.  Each point indicates the
relationship between \NCO\ and \NHt, and the color identifies its
value of the \X\ factor, calculated from Equation \ref{Xfac}.

For the Milky Way MC model (n300, Fig. \ref{Xfig}a), in $>$90\% of
positions the \X\ factor is within the range $9\times 10^{19} -
5\times 10^{20}$ \Xunits.  Figures \ref{pdfpans}a-b show that the
quantities \NHt, \NCO, and \W\ all have similar (Gaussian-like)
shapes, which ultimately result in a limited range in the distribution
of the \X\ factor.  The range in the derived \X\ factor for this model
is in agreement with estimates from observations of Milky Way clouds
\citep[e.g][]{Solomonetal87, Young&Scoville91, Dameetal01}.  This
agreement gives us confidence that the simulations, including the
chemical networks as well as the radiative transfer calculations, are
sufficiently reliable and can be utilized to probe conditions
different from the Milky Way.

The distribution of the \X\ factor increases substantially when the
observed \CO\ emission is poorly correlated with the CO column
densities, as in Model n300-Z03 (Fig. \ref{Xfig}b).  In this model,
the \X\ factor distribution spans a wide range: only $<$ 25\%
have values in the range $9\times10^{19} - 5\times10^{20}$, and
$>$30\% have values $> 10^{22}$ \Xunits.

Figure \ref{compfigs}c and d shows \NHt\ and the \X\ factor for each
line-of-sight of Model n300-Z03.  In the highest density regions such
as the large complex at the top right-hand side of CO or \Ht\ column
density maps (Figs. \ref{compfigs}a and c), the \X\ factor has a
limited range, within one order of magnitude ($10^{20} - 10^{21}$
\Xunits).  In those regions, though \NCO\ can vary by 2 or more orders
of magnitude, \W\ only varies by $\approx$ 1 order of magnitude.
Since the \X\ factor directly depends on \W\ and only indirectly on
\NCO, the \X\ factor also only falls into a limited range.

Positions with the largest \X\ factors correspond to the lowest
\NHt\ regions, as well as low \NCO\ and \W\ regions.  These are the
regions where CO is most affected by photodissociation.  Since the
amount of photodissociation depends on the the ``effective'' column
density in each location of the 3D simulation volume, regions with
similar \NHt\ can have very different \NCO\ values (see also Paper I
and II), as evident in Figure \ref{Xfig}b: at low to intermediate
\Ht\ densities $10^{21}$\aplt \NHt \aplt\ $10^{22}$ \cmsq, there is a
wide range of \NCO\ for a given \NHt.  Since the \X\ factor
(indirectly) depends on \NCO, at such densities the \X\ factor also
takes on a wide range.  For instance, at \NHt\ $=5\times10^{21}$
\cmsq, the \X\ factor varies from $\sim 10^{20}$ to $10^{23}$ \Xunits.
Evidently, the \X\ factor can have a wide distribution within a MC,
even for regions with identical molecular column densities.  This is a
consequence of the combination of a large distribution of \NCO\ for a
given \NHt, as well as the lack of simple correlation between \W\ and
\NCO\ due to the optically thick nature of \CO.

Figure \ref{Xfig}d shows that there can be a wide range in the
\X\ factor even in very low density regions.  For this model ($n_0$ =
100 \cmsq\ and $Z = Z_\odot$), much of the gas has \NHt \aplt
10$^{21}$ \cmsq.  Unlike Model n300 (in Fig. \ref{Xfig}a), there is a
very wide distribution in the \X\ factor in the range $10^{20}$ \aplt
\NHt \aplt $10^{21}$ \cmsq.  This model also differs from Model
n300-Z03 (in Fig. \ref{Xfig}b), showing a much larger distribution in
\NCO\ and the \X\ factor for a given \NHt\ at \NHt
\aplt\ $2\times10^{21}$ \cmsq.  The multiple peaks in the \X\ factor
distributions can be understood from the \W\ and \NCO\ distributions
shown in Figure \ref{pdfpans}h.  Model n100 has a large distribution
of \NCO; the \W\ distribution is similar, with multiple peaks (which
are not aligned with the \NCO\ peaks), contributing to the wide range
and multiple peaks of the \X\ factor shown in Figure \ref{Xfig}d.

At very high densities, more self shielding allows for CO formation to
be more effective. As Figure \ref{Xfig}c shows for Model n1000,
\NCO\ is a much better tracer of \NHt.  However, though \NHt\ for this
model reaches values 3 times greater than Model n300
(Fig. \ref{Xfig}a), the distribution in the \X\ factors of both models
is very similar.  In Model n1000, the \X\ factor increases with
increasing (CO or \Ht) density, a trend opposite to that apparent for
the other models shown in Figure \ref{Xfig}.

\begin{figure*}
\includegraphics[width=175mm]{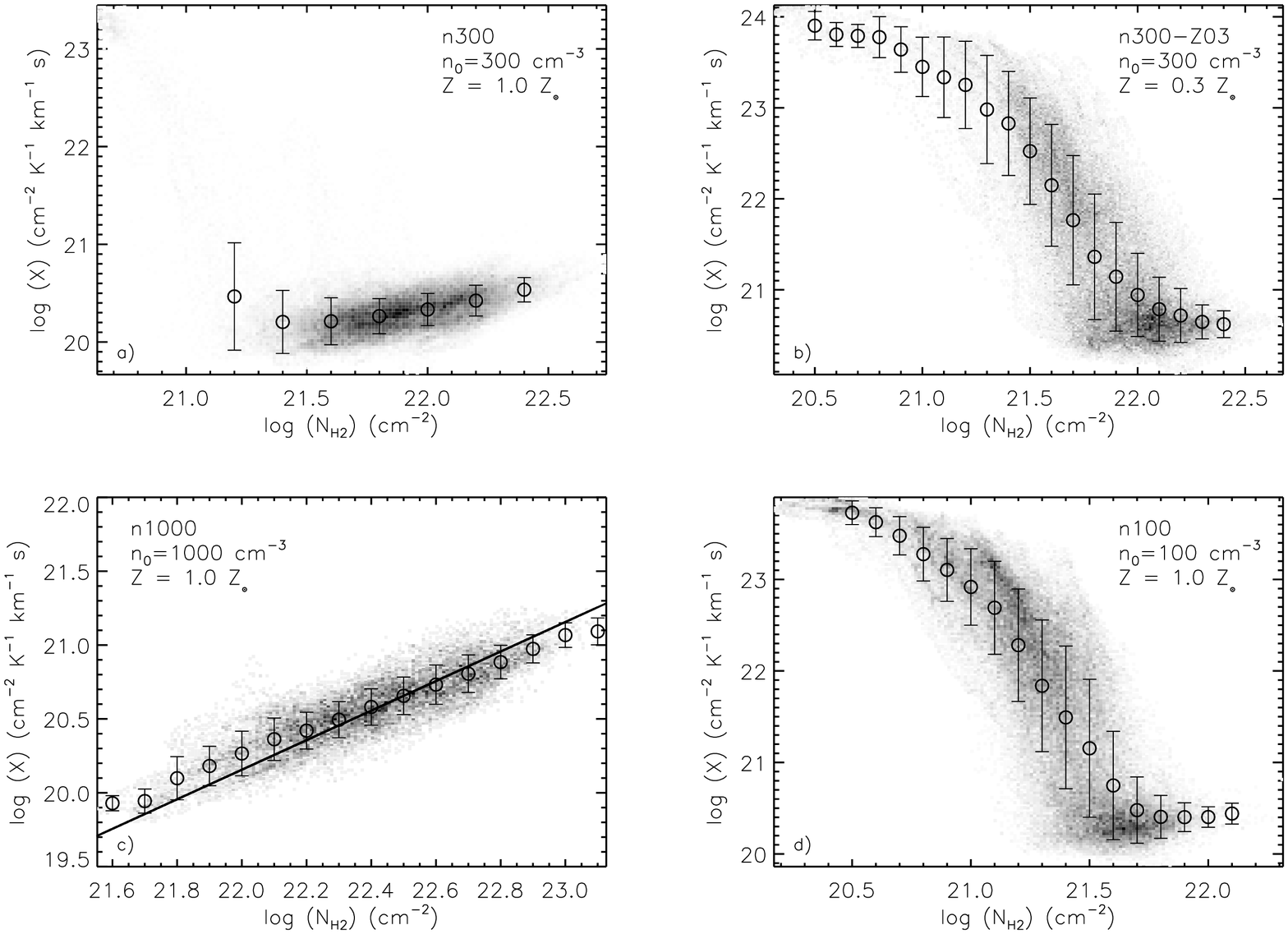}
\caption{\X\ factor plotted against \NHt\ for the 4 models shown in
  Figure \ref{Xfig}.  The (grey scale) color indicates the frequency
  of a given \NHt\ and \X\ factor pair, where darker points indicate
  higher frequency.  Black circles show average \X\ values in
  \NHt\ bins, and the error bars indicate 1$\sigma$ deviations.  The
  line in bottom left panel shows \X\ $\propto$ \NHt\ relationship.}
\label{XvsAv}
\end{figure*}

The correlation between the \X\ factor and \NHt\ is shown in Figure
\ref{XvsAv}.  Model n300 shows that at \NHt \apgt $10^{21}$ \cmsq, the
\X\ factor is $\sim$ constant (with a mean value $2\times 10^{20}$
\Xunits).  Only at the highest densities (log(\NHt) \apgt 22.0) do the
effects of saturation become clearly evident.  On the other hand,
Model n1000 (shown in Fig. \ref{XvsAv}c) shows an increase in the
\X\ factor at all \NHt.  The line shows the \X\ = \NHt/(70 \Wunits)
relationship, which is simply the \Ht\ column density divided by the
mean CO intensity from this model.  This line is a very good fit at
log(\NHt) \apgt 22, indicating that at most densities the CO line
emission is completely saturated.

The models with lower metallicity and density (Figs. \ref{XvsAv}b and
\ref{XvsAv}d) show a decreasing \X\ factor with increasing density (as
can also be seen in Fig. \ref{Xfig}).  In this regime, an increase in
the molecular density is associated with a greater increase in CO line
intensity, resulting in a decrease in the \X\ factor.  Such a trend is
also present at the lowest densities of Model n300
(Fig. \ref{XvsAv}a).

Given that \W\ can have rather different distributions than \NCO, it
comes as no surprise that the \X\ factor does not neatly follow from
\NCO.  For clouds with high CO abundances, such as models n300 and
n1000, \NCO\ is well correlated with \NHt, resulting in a limited
range in the distribution of the \X\ factor.  We have seen that CO
line saturation may not be discernible solely from the \X\ factor
distributions, as evident from the similarity of the \X\ factor
(normal) distributions of Model n300 (with saturation occurring only
at the highest densities) and n1000 (which is close to fully
saturated).  The relationship between \X\ and
\NHt\ (Fig. \ref{XvsAv}), however, clearly demonstrates the saturation
of the CO line.  For the models with lower density or metallicity,
where the \W\ - \NHt\ correlation is rather complex, the \X\ factor
can have a large distribution, and can even have multiple peaks in its
PDF (such as Model n100, Fig \ref{Xfig}d).

\section{Discussion and Comparison to Observations}\label{discsec}

\subsection{The Column Density Distributions: Can CO Observations Reveal Log-Normal PDFs? \label{logdisc}}

Observed morphological characteristics, such as the density
distribution, may be signatures of the internal dynamics of a cloud.
Numerous simulations have shown that the (3D) volume density PDF is
log-normal for non self-gravitating supersonic turbulent systems
\citep[see][and references therein]{McKee&Ostriker07}, which may be
translated into log-normal PDFs of projected (2D) column density
\citep{Ostrikeretal01, Bruntetal10a, Bruntetal10b}.  Including gas
self-gravity, which may counteract the dispersal due to turbulence so
that some overdense regions may persist as bound objects, results in a
PDF with a high density tail \citep{Klessen00}.  These signature high
density tails have been found in column density PDFs obtained from
observations of evolved clouds \citep{Kainulainenetal09}.  In general,
however, directly correlating features of the PDF to physical
processes in the simulation has proven to be rather difficult
\citep{Klessen00, Federrathetal08, Federrathetal10}.

Understanding the relationship between CO and the underlying density
distribution is essential if CO observations are employed as
(molecular) mass tracers of MCs.  Usually, density distributions are
derived from $^{13}$CO intensities, since $^{13}$CO is optically thin
(relative to $^{12}$CO).  In our analysis of $^{12}$CO emission, we
find that \W\ does not faithfully trace \NCO\ even in some low density
regions.  This is suggestive that $^{13}$CO intensities may also not
be an ideal tracer of the underlying CO distribution.  Though we
analyze the density distribution traced by $^{12}$CO emission in this
work, the suitability of $^{13}$CO as a column density tracer remains
an open question that we plan on investigating in future work.

In practice, the shape of observed density PDFs strongly depends on
the tracer observed.  For instance, \citet{Goodmanetal09b} found that
dust observations of the Galactic MC Perseus show log-normal column
density PDFs, and may provide reasonable estimates of the total mass.
Gas ($^{13}$CO) based measures result in PDFs with significantly
different shapes, though a log-normal fit can loosely describe the
distribution.  In this section, we compare the PDFs of observed CO
intensities with those of the intrinsic \Ht\ and CO column density
(see Paper I for a discussion on the volume density PDFs).

We have found that the observed \W\ PDF can vary significantly from
the column density PDFs, \NCO\ or \NHt, even though \NHt\ (and \Ntot)
for all models can be well described as log-normal distributions, as
shown in Figure \ref{pdfpans}.  Only models n300 and n1000 seemingly
have \NCO\ and \W\ PDFs that can be well described as log-normals.
For the PDFs of those models (Fig \ref{pdfpans}a,b,e,f), we have
overplotted best fit log-normals, with functional form
\begin{equation}
{\rm PDF} \propto e^{-(\log\, x-\log\, x_0)/2\sigma^2}, 
\label{bestfitlog}
\end{equation}
where $x$ is \Ntot, \NHt, \NCO, or \W, $\sigma^2$ is the variance, and
$x_0$ denotes the mean value.  The distributions shown in Figure
\ref{pdfpans} show that the offsets in the log-normals do not
coincide.  Of course, this discrepancy is partly due to our choice of
the axis ranges of \NCO\ and \W.  Thus, a comparison of $\sigma$,
which is a measure of the width of the log-normal, is the parameter
best suited for direct comparison between the PDFs.

For both models n300 and n1000, the \Ntot\ and \NHt\ PDFs have
$\sigma=$ 0.3.  For Model n300, $\sigma=$ 0.5 and 0.2 for the
\NCO\ and \W\ PDFs, respectively.  The best fits give $\sigma=$ 0.3
and 0.1 for the \NCO\ and \W\ PDFs, respectively, for Model n1000.  As
can be seen from the PDFs themselves, the best fits \NCO\ and
\W\ log-normals are rather different for the two models.  Such
discrepancies are not surprising, given that the other models exhibit
PDFs of \W\ and \NCO\ with vastly different forms.

That the \W\ and \NCO\ PDFs from Model n1000 are both log-normal,
albeit with different characteristics, may be somewhat unexpected,
given that this model is clearly affected by saturation
(Figs. \ref{wvsav} and \ref{XvsAv}).  With line saturation, one would
expect the \W\ PDF to be asymmetric, with a prominent peak towards
large intensities, and a sharp drop off beyond the peak - the
``piled-up'' effect discussed in Section \ref{intensitysec}.  At any
given extinction from Model n1000, even though much of the line
emission from the highest density gas is saturated, there is a
sufficiently large range of \W\ such that there is no sharp drop off
in the \W\ PDF beyond the peak.  This demonstrates that line
saturation may be significant even when the PDF of the observed
intensities does not show a sharp decrease.  For such scenarios, it is
only when inspecting \W\ with \Ntot\ (Fig. \ref{wvsav}) or \X\ with
\NHt\ (Fig. \ref{XvsAv}), for which independent estimates of \Ntot\ or
\NHt\ would be required, that the presence of saturation can be
inferred.

In general, we find that the \W\ PDFs are not log-normal, even though
the underlying gas column densities are.  One reason for the
difference is that the \NCO\ PDFs themselves are not log-normal.
Further, in most of the models, the \W\ PDF itself does not neatly
follow that of \NCO.  Since CO traces the dense gas, \NCO\ is more
intermittent than the underlying gas distribution.  Due to the effects
of saturation, this intermittency is not captured by \W.  In the two
models with log-normal \W\ and \NCO\ PDFs, the variances between the
distributions are nevertheless different.  Taken together, we conclude
that the distribution of observed CO intensities is not an accurate
measure of the underlying distribution of molecular gas.  This
conclusion is affirmed by the observational analysis of
\citet{Goodmanetal09b}, who found significant differences in the
$^{13}$CO and dust-based distributions.

\subsection{Globally Averaged X Factor}

We have shown that, in general, the CO integrated intensity does not
linearly trace \NHt, even within individual MCs, so that the
\X\ factor is not constant.  The intrinsic cloud properties, such as
density, metallicity, and background UV radiation field, all influence
how \W\ is related to \NHt.  In the previous sections, our analysis
has dealt with all positions in the 2D ``observed'' maps, which are
idealized synthetic maps with high spatial resolution of $\sim$ 0.1 pc
(the size of the grid zone in the MHD simulations).  However, in
practice the derived \X\ factor is likely to be an average over a
range in extinction values, especially in observations of
extragalactic clouds where the resolution may be comparable to the
cloud size.  In order for a more direct comparison with these types of
integrated observations, we now shift our focus to the discussion of
average quantities in the synthetic maps.

\begin{figure*}
\includegraphics[width=100mm]{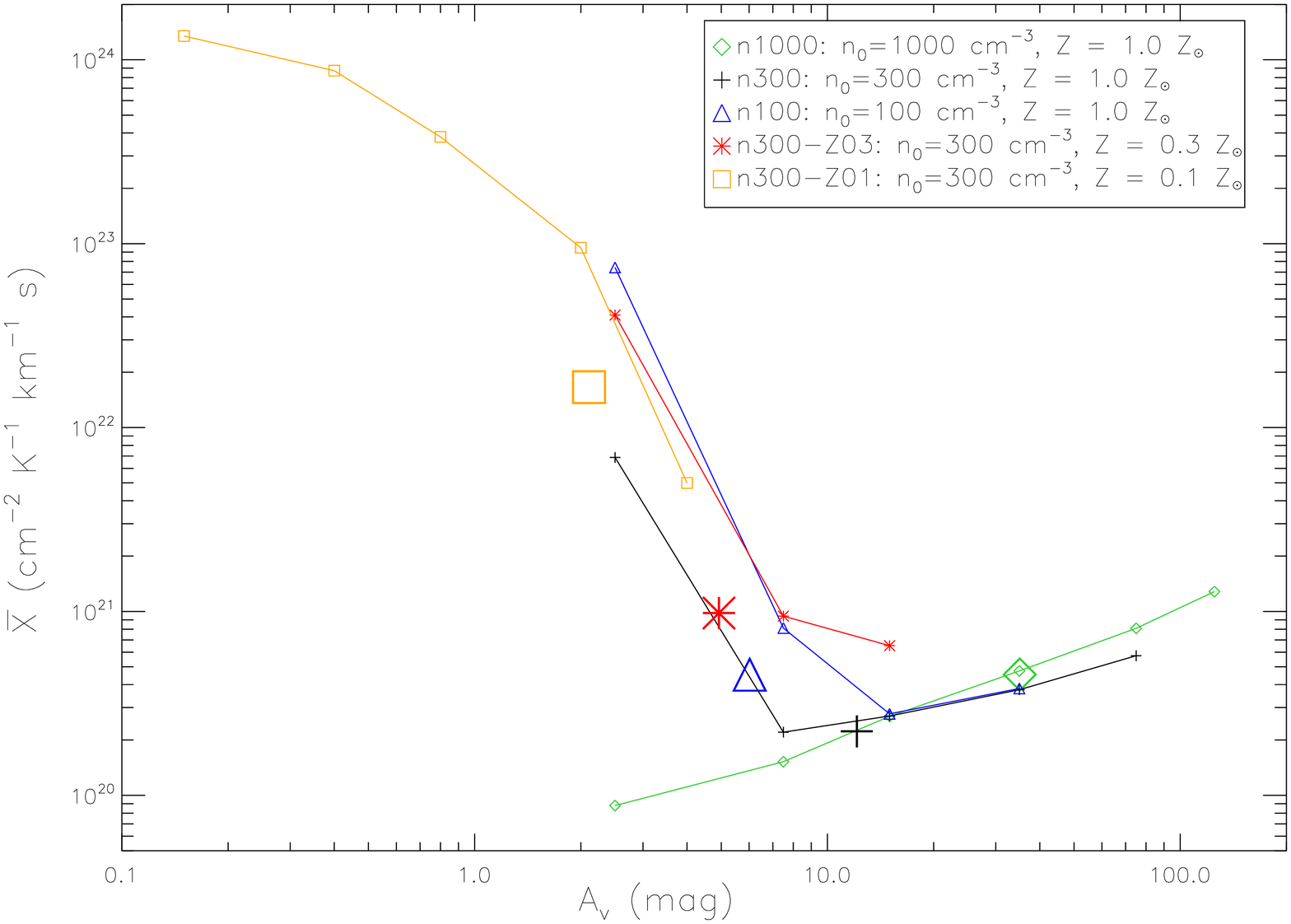}
\caption{Mean \X\ factor in \Av\ bins for 5 models.  The \X\ factor is
  averaged in different \Av\ bins.  The value of $\overline{X}$ is
  plotted on the midpoint value of \Av\ of each bin.  Each model is
  identified by different colors and symbols (and labeled in the
  legend).  The large symbols shows the global (emission weighted)
  mean \X\ factor and mean \Av\ from each model.}
\label{meanX}
\end{figure*}

Figure \ref{meanX} shows the mean \X\ factor in different \Av\ bins
for the four models shown in Figures \ref{Xfig}-\ref{XvsAv}, as well
as Model n300-Z01.  The emission-weighted mean \X\ and \Av\ for each
model is also shown (by the large symbols).  A number of features of
the averaged \X\ factor are similar to those already discussed: a
decrease in \X\ with increasing extinction for models with low
densities or metallicities, the opposite trend for Model n1000 (and to
some extent Model n300) due to line saturation, and only a small
variation in \X\ at high densities for Model n300.  Qualitatively, the
mean values of the \X\ factor binned with extinction resemble the
extinction averaged quantities presented in Paper II (see their
Fig. 8).

One clear difference is that values of the \X\ factor averaged over a
range in extinctions for a given model do not span as wide a range as
when all positions are considered (compare, e.g. to
Figs. \ref{Xfig}-\ref{XvsAv}).  Only when comparing all the models at
the low extinction \Av \aplt\ 5 does the large extent in the
\X\ factor ($10^{20} < $ \X\ $< 10^{24}$ \Xunits) become clear.

The striking feature of Figure \ref{meanX} is the limited variation in
the \X\ factor at large extinctions, both for a given model as well as
between models.  At extinctions \apgt7, \X\ falls in a narrow range
$\sim 10^{20}$ \aplt \X \aplt $10^{21}$ \Xunits\ for all
models.\footnote{This threshold extinction beyond which the \X\ factor
  is constant is larger by a factor of $\sim$2 than that found in
  Paper II.  The discrepancy arises because the analysis in Paper II
  relied on simplified assumptions to compute the CO intensities,
  whereas here we use more accurate radiative transfer calculations.}
This range is similar to the \X\ factor values derived for the Galaxy
\citep{Solomonetal87,Young&Scoville91,Dameetal01}.  The emission
weighted mean \X\ and \Av\ values (large symbols) also portray a near
constant \Xgal\ value between the MCs, with the exception of the very
low metallicity MC.  Thus, only considering CO bright regions,
corresponding to high extinctions, may not expose any real variation
in the \NHt/\W\ ratio, as discussed in Paper II.

Observations of the Small Magellanic Cloud (SMC), which has lower than
Galactic metallicity, has resulted in high \X\ factor estimates
\citep[][and references therein]{ Israel97, Leroyetal09}.  Averaged
over a whole cloud in the SMC, \citet{Leroyetal09} found that the
\X\ factor varies by \apgt1 order of magnitude, whereas the \X\ factor
computed at the CO peaks is similar to the Galactic value.  Our
analysis of CO emission from models with lower metallicity, especially
Model n300-Z03, which has a metallicity similar to that of SMC
\citep{Vermeij&vanderHulst02}, quantitatively reproduces the estimated
\X\ factor, both the larger value due to diffuse gas and the lower
Galactic value found in the densest regions.

\subsection{\X\ Factor in Dense and Diffuse Gas}

In our discussion of the Galactic \X\ factor, we have focused on
observations of molecular clouds \citep[e.g.][]{Young&Scoville91,
  Solomonetal87}.  However, CO has also been detected in the diffuse
ISM \citep[e.g.][]{Polketal88,Burghetal07}, and the \X\ factor for
these diffuse lines of sight is found to be consistent with the
standard galactic value of \Xgal\ \citep[few $\times10^{20}$
  \Xunits,][]{Lisztetal10}.  Though we consider models of individual
(or large volumes within) MCs, we investigate whether regions with
different densities are comparable with the observations of different
phases of the ISM, focusing on the recent discussion on the uniformity
of the \X\ factor in \citet{Lisztetal10}, hereafter LPL10.

LPL10 find that diffuse gas, with molecular column densities
\NHt\ $\sim$ 10$^{20}$--10$^{21}$ \cmsq, and dense Galactic clouds,
characterized by \NHt\ $\sim$ 10$^{23}$ \cmsq, are always found to
have \X\ factor values that are $\sim$\Xgal.  Since the \X\ factor can
be written as
\begin{equation}
X=\left( \frac{N_{\rm CO}}{W} \right) \left( \frac{N_{\rm H_2}}{N_{\rm CO}} \right) = \frac{N_{\rm CO}}{W} \frac{1}{f_{\rm CO}}, 
\label{Xchem}
\end{equation}
LPL10 attribute the constancy of the \X\ factor to the increase in the
specific brightness, $W/N_{\rm CO}$, in low density, warm gas, by a
factor comparable to the decrease in $f_{\rm CO} = N_{\rm CO}/N_{\rm
  H_2}$.\footnote{LPL10 use the symbol $X_{\rm CO}$, whereas we will
  use $f_{\rm CO}$ to avoid confusion with the \X\ factor.}  For
diffuse gas, LPL10 use an estimate of $f_{\rm CO} \approx 3\times10^{-6}$, 
based on \citet{Burghetal07}, while for dense gas, they adopt
$f_{\rm CO} \approx 10^{-4}$.  These values correspond
to $W/N_{\rm CO} \approx 10^{-15}$ and $3\times10^{-17}$
\Kkms\ cm$^2$, in diffuse and dense gas, respectively, for a constant
\X\ factor $=3\times10^{20}$ \Xunits.

As discussed, we have found that the Milky Way model n300 does indeed
result in \X$\approx$\Xgal\ in most lines of sight.  In this model,
there is only a very small fraction of the simulated volume with the
lowest \Ht\ column densities \NHt \aplt $10^{21}$ \cmsq\ (see
Fig. \ref{Xfig}a).  Thus, for comparison with LPL10 we will use model
n300-Z03, which has many lines of sight with both dense (\NHt
$\sim10^{22}$) and diffuse gas (\NHt $\sim5\times10^{20}$).  Maps of
the various relevant quantities for this model are presented in Figure
\ref{compfigs}.  Though this model has lower metallicity ($Z$ =
0.3\zsun) than the observations of LPL ($Z$=\zsun), a comparison of
the observations with Model n100, (or the high density gas in Model
n300) bear similar results as those provided by Model n300-Z03.

We identify dense gas as regions with log($W$) $>$ 1.5, marked by
dashed contours in Fig. \ref{compfigs}, and compute $W/N_{\rm CO} \in
4\times10^{-17} - 8\times10^{-16}$ \Kkms\ cm$^2$.  The lower and upper
limits have corresponding $f_{\rm CO}$ values of $5\times10^{-5}$ and
$5\times10^{-6}$, respectively.  The resulting value of the \X\ factor
is $\sim 1-6 \times 10^{20}$ \Xunits.  These values of $W/N_{\rm CO}$,
$f_{\rm CO}$, and \X, reasonably reproduce the values determined by
LPL10 for dense clouds, with an \X\ factor which only differs by at
most a factor of 3.

The low density regions of Model n300-Z03, marked roughly by the
log($W$) $<$ -1 contour in Fig. \ref{compfigs}, have $W/N_{\rm CO}
\sim 10^{-15}$ \Kkms\ cm$^2$, and $f_{\rm CO} \sim 6\times10^{-10}$,
resulting in an \X\ factor $\sim 10^{24}$ \Xunits.  This \X\ factor
differs from the LPL10 estimated value from observations of diffuse
gas by up to 4 orders of magnitude.  The discrepancy can be almost
fully attributed to the difference in $f_{\rm CO}$ between our models
and the LPL10 observations; the values of $W/N_{\rm CO}$ of our models
and that determined by the LPL10 analysis only differ by a factor of
$\sim$ 2 - 10.  Table \ref{LPLcomp} summarizes the comparison of
$W/N_{\rm CO}$, $f_{\rm CO}$, and \X\ between Model n300-Z03 and LPL10
in the dense and diffuse gas.

\begin{table*}
   \centering
 \begin{minipage}{140mm}
  \caption{Comparison of Model n300-Z03 and \citet{Lisztetal10} Observations}
  \begin{tabular}{ccccc}
  \hline
  \hline
  Gas Phase & \NHt\ (\cmsq)  & $f_{\rm CO}$ & $W/N_{\rm CO}$  & \X\ (\Xunits) \\
  \hline

Observed Dense Gas (LPL10)  & $\sim10^{23}$ & $\sim10^{-4}$ & $\sim3\times10^{-17}$ & $\sim3\times10^{20}$ \\
Dense Gas in Simulation (n300-Z03) & $\sim2-4\times10^{22}$ & $\sim0.5-5\times10^{-5}$ & $\sim0.4-8\times10^{-16}$ & $\sim 1-6 \times 10^{20}$ \\
\\
Observed Diffuse Gas (LPL10)  & $\sim10^{20} - 10^{21} $ & $\sim3\times10^{-6}$ & $\sim10^{-15}$ & $\sim3\times10^{20}$ \\
Low Density Gas in Simulation (n300-Z03) & $\sim2-7\times10^{20}$ & $\sim6\times10^{-10}$ & $\sim10^{-15}$ & $\sim 10^{24}$  \\
\hline
\end{tabular}
\label{LPLcomp}
\end{minipage}
\end{table*}

The comparison of the characteristics of the dense regions has
quantitatively affirmed some level of accuracy of the chemical (and
MHD) modeling, as well as the radiative transfer calculations.
Additionally, we can draw some insights from the discrepancy we find
in the characteristics of the diffuse regions: the larger difference
in $f_{\rm CO}$ in our models compared to the LPL10 observations of
tenuous gas, rather than $W/N_{\rm CO}$, indicates that the chemical
processes at work in the low density regions within our molecular
clouds are very different from the chemistry of diffuse Milky Way gas.
That $f_{\rm CO}$ is significantly lower in the low density regions of
the model suggests that either CO forms at a greater rate in the
diffuse ISM (relative to the rate in the model), or that CO is
efficiently transported into the diffuse ISM (and has sufficient
shielding against photodissociation).  The physical processes at work
in our models, turbulence, magnetic fields, or simply even the uniform
metallicities (or initial densities), may be significantly different
in the tenuous ISM.  In particular, non-thermal chemistry in C-shocks
or turbulent vortices may play an important role in the chemistry of
the diffuse gas \citep[see
  e.g.][]{Federmanetal96,Shefferetal08,Godardetal09}.  Alternatively,
large scale galactic processes not considered in our model may be
required to accurately model the chemistry in diffuse molecular gas,
such as galactic rotation, the gravitational field due to the disk,
and/or large scale magnetic fields.  We hope to pursue studies
involving such large scale dynamics and chemistry in the future.

\subsection{Future Work: \X\ factors as Mass Estimators for CO Bright Cores?}

In observational investigations where the \X\ factor is derived solely
from CO observations, the CO linewidths are used to derive masses
through the virial theorem.  If the bulk of the gas is in molecular
form, then \W\ is employed along with the CO linewidth derived
\NHt\ to estimate the \X\ factor via Equation \ref{Xfac}.  The key
assumption here is that the observed clouds are in virial equilibrium
\citep[e.g.][]{Young&Scoville91, Solomonetal87}, which is the
interpretation of the observed power-law linewidth $\sigma_v$ size $R$
relationship $\sigma_v \propto R^{0.5}$ \citep[e.g.][]{Larson81,
  Solomonetal87}.  Measurements of the molecular mass through
independent observations, such as dust-based emission or extinction,
can also provide estimates of \NHt.  These can be combined with CO
observations to calculate the \X\ factor.  As discussed in Section
\ref{introsec}, a discrepancy in the derived \X\ factor based on
virialized clouds or based on independent \Ht\ mass estimates is found
in low metallicity systems, such as the SMC \citep[][]{Bolattoetal08,
  Rubioetal04, Israel97, Leroyetal07, Leroyetal09}.

The velocity dispersions in our models follow the observed $\sigma_v
\propto R^{0.5}$ relationship (Shetty et al. In preparation), as found
in numerous similar turbulent cloud models \citep[e.g.][]{Klessen00,
  Ostrikeretal01, Federrathetal10}.  We have calculated the
\X\ factors in model MCs assuming independent knowledge of \NHt, so no
assumption of a linewidth-size scaling, or of virialized clouds (or
cores), is necessary.  In our analysis of total column densities and
integrated intensities, we find that \X\ factor is primarily
controlled by the CO abundance, and thus ultimately by cloud
properties most responsible for CO formation: the metallicity,
density, or background UV radiation field.  This variation in the
\X\ factor is most drastic - upto several orders of magnitude - in
models with metallicities, densities, or UV radiation fields different
from the Milky Way.  These results are in agreement with models
showing an increased \X\ factor for low metallicity systems
\citep{Maloney&Black88, Israel97}.

The discrepancy between \X\ factor estimates based on virialized
systems, and those based on independent \NHt\ measures can be
explained by the selective photodissociation of CO in low density
regions.  The CO bright objects thus only trace highest density gas,
which is surrounded by lower density molecular material, with little
or no CO emission \citep[][Paper II, Molina et al. in
  prep.]{Maloney&Black88,Bolattoetal08, Grenieretal05, Wolfireetal10}.
In low metallicity or low density systems, we only find the \X\ factor
to be constant at the very highest densities (Figs. \ref{XvsAv}b \&
\ref{XvsAv}d), though the vast majority of observed points clearly
show a gradient in the \X\ factor with density.  For the high density
Model n1000 (Fig \ref{XvsAv}c), the \X\ factor increases with
increasing density due to CO line saturation (Fig. \ref{wvsav}c),
though the range of \X\ is only \aplt\ 2 orders of magnitude.

The idea that CO observations accurately provide mass estimates, with
\X $\approx$\Xgal, of virialized clouds needs to be affirmed
quantitatively.  Previous efforts to address this issue have usually
involved static models.  \citet{Kutner&Leung85} constructed clouds
models with microturbulent velocities consistent with the observed
linewidth-size relationship.  They found that the \X\ factor still
strongly depends on the temperature, as well as the density and CO
abundance.  Using photodissociation region models,
\citet{Wolfireetal93} found that microturbulence cannot reproduce the
observed CO line profiles.  The CO intensity, and hence the
\X\ factor, is sensitive to other parameters responsible for CO
formation.  With the radiative transfer calculations performed on the
3D MHD simulations including chemistry discussed here, we can further
investigate how various turbulent velocity fields influence CO
emission, as well as the cloud characteristics that affect the
\X\ factor in a range of environments.

Such an analysis would address the following issues: How well do CO
peaks trace coherent objects in spectral (PPV) cubes?  Identifying
coherent objects from spectral cubes has proved to be quite
challenging, as a consequence of line-of-sight projection
\citep{Adler&Roberts92,Pichardoetal00}.  How well do CO linewidths
encode information on the dynamics of the cloud?  Projection effects
may also skew the linewidth - size power law scalings
\citep{Ballesteros-Paredesetal99,Ballesteros-Paredes&MacLow02,Shettyetal10}.
If a CO bright core can be accurately identified, is the observed
linewidth representative of the intrinsic velocity dispersion?  To
what extent does turbulence affect the CO emission, and thus the
\X\ factor?  These issues will all be visited and discussed in a
forthcoming paper.

\section{Summary} \label{sumsec}

We have performed radiative transfer calculations to investigate the
nature of $^{12}$CO (J=1-0) emission from simulations of molecular
clouds (MCs).  The MCs are modeled through hydrodynamic simulations of
a turbulent, magnetized, non self-gravitating gaseous medium, along
with a treatment of chemistry to track the formation of \Ht\ and CO
(Paper I and Paper II).  As part of the radiative transfer
calculations, we use the Sobolev (LVG) method to solve for the CO
level populations.  We analyze the probability distribution functions,
and \X\ factor properties, using the velocity integrated CO
intensities \W, along with the column density of CO and \Ht, \NCO\ and
\NHt, respectively.  Our main findings are:

1) In all models, \W\ increases with increasing total hydrogen column density
\Ntot\ (or extinction \Av).  However, for the Milky Way and high
density models (n300 and n1000), which have the highest CO abundances,
we find a threshold in \W $\approx$65 \Kkms\ at high extinction due to
saturation of the CO line.

2) All models have log-normal \Ntot\ and \NHt\ distributions.  In
general, however, the \W\ and \NCO\ PDFs are not log-normal.  Further,
since CO is optically thick, the \W\ PDFs do not have similar shapes
to the corresponding \NCO\ PDFs.

3) In some models for which the CO line is saturated, the peak in
\NCO\ is offset towards higher densities than the peak in \W, though
the two PDFs seem to be correlated at low densities.  However, such a
``piled-up'' \W\ PDF does not necessarily arise in clouds with
saturated CO emission, especially those with a limited range in
densities (as for the high density model n1000,
Figs. \ref{wvsav}-\ref{pdfpans}). Independent measurements of
\Ntot\ or \NHt, along with \W, are needed to unambiguously identify
saturation.

4) The \X\ factor is not constant within individual molecular clouds,
and in models with low CO fractions, can vary by up to 4 orders of
magnitude.  The low density regions have the highest \X\ factor, in
agreement with previous modeling efforts.

5) In most simulations, the averaged \X\ factor is found to be similar
to \Xgal\ ($ \sim2 \times 10^{20}$ \Xunits), masking the variation of
\X\ with density within clouds.  In clouds with low CO abundances
relative to the Galaxy (such as the model MC in a dwarf galaxy or the
LMC/SMC), the densest regions have \X\ $\sim$ \Xgal.

6) Emission weighted averaged \X\ factors from all models provide
values $\approx$\Xgal\, with the exception of the very low metallicity
MC (n300-Z01).  Similarly, at extinctions \Av\ \apgt7, the \X\ factor
for always falls in a narrow range $\sim 10^{20}$ \aplt \X \aplt
$10^{21}$ \Xunits.  As discussed in Paper II, observations targeting
CO bright regions may be unable to detect real variations in the
\X\ factor.

7) In the low density gas within the MC models, the CO fraction
$f_{CO}$ is found to be $\sim$ three orders of magnitude lower than
$f_{CO}$ measured by \citet{Burghetal07} in diffuse gas.  That
\NHt\ and \W/\NCO\ are comparable between the model and the
observations suggest that the dynamics and chemical evolution of
diffuse gas in our molecular cloud model are rather different from
that in the large scale diffuse gas of the Galaxy.

8) We do not assume that clouds or cores are in virial equilibrium,
and the \X\ factor variations we find are in general agreement with
observational analyses employing independent measures of total
molecular column densities.  In a follow up investigation, we will use
the spectral information of our models to assess the \X\ factor when
virial equilibrium is assumed; we will then be able to address the
\X\ factor discrepancy found in observational works when assuming
virialized CO clouds or when using independent molecular mass
estimates.

\section*{Acknowledgements}
We are grateful to Eve Ostriker, Alyssa Goodman, Jaime Pineda,
Mordecai Mac Low, Frank Bigiel, Andrew Harris, and Christoph Federrath
for stimulating discussions regarding CO emission and the \X\ factor.
We also thank an anonymous referee for useful
comments. R.S.K. acknowledges financial support from the
Landesstiftung Baden-W\"urrtemberg via their program International
Collaboration II (grant P-LS-SPII/18) and from the German
Bundesministerium f\"ur Bildung und Forschung via the ASTRONET project
STAR FORMAT (grant 05A09VHA). R.S.K. furthermore acknowledges
subsidies from the DFG under grants no. KL1358/1, KL1358/4, KL1358/5,
KL1358/10, and KL1358/11, as well as from a Frontier grant of
Heidelberg University sponsored by the German Excellence
Initiative. This work was supported in part by the U.S. Department of
Energy contract no. DE-AC-02-76SF00515. R.S.K. also thanks the Kavli
Institute for Particle Astrophysics and Cosmology at Stanford
University and the Department of Astronomy and Astrophysics at the
University of California at Santa Cruz for their warm hospitality
during a sabbatical stay in spring 2010.

\bibliography{citations}

\label{lastpage}
\end{document}